\newif\ifpreprint

\preprintfalse 

\ifpreprint
\documentclass[aip,jcp,amsmath,amssymb,preprint]{revtex4-1}
\else
\documentclass[aip,jcp,amsmath,amssymb,reprint]{revtex4-1}
\fi

\usepackage{graphicx}		
\usepackage{amssymb}
\usepackage{amsmath}
\usepackage{braket}
\usepackage{xcolor}
\usepackage{xspace}
\usepackage{ifthen}

\newcommand{\sqrot}[2]{\hat{\kappa}^{#1}_{#2}}

\newcommand{\sqop}[2]{\hat{a}^{#1}_{#2}}
\newcommand{\cop}[1]{\hat{a}_{#1}^{\dagger}}
\newcommand{\aop}[1]{\hat{a}_{#1}^{\,}}

\newcommand{\uccf}[0]{UCC path equation\xspace}
\newcommand{\flow}[0]{path\xspace}
\newcommand{\theoryname}[0]{disentangled UCC\xspace}
\newcommand{\theorypre}[0]{disentangled\xspace}

\definecolor{goodorange}{RGB}{225,125,0}
\definecolor{goodgreen}{RGB}{5,130,5}
\definecolor{goodred}{RGB}{220,50,25}
\newcommand{\note}[2]{
\ifthenelse{\equal{#1}{F}}{
\colorbox{goodorange}{\textcolor{white}{\footnotesize \fontfamily{phv}\selectfont #1}}
    \textcolor{goodorange}{{\footnotesize \fontfamily{phv}\selectfont #2}}\xspace
}{}
\ifthenelse{\equal{#1}{G}}{
\colorbox{goodred}{\textcolor{white}{\footnotesize \fontfamily{phv}\selectfont #1}}
    \textcolor{goodred}{{\footnotesize \fontfamily{phv}\selectfont #2}}\xspace
}{}
\ifthenelse{\equal{#1}{N}}{
\colorbox{goodgreen}{\textcolor{white}{\footnotesize \fontfamily{phv}\selectfont #1}}
    \textcolor{goodgreen}{{\footnotesize \fontfamily{phv}\selectfont #2}}\xspace
}{}
}

\begin{document}
\include{main_text}
\title{Exact Parameterization of Fermionic Wave Functions via Unitary Coupled Cluster Theory}

\author{Francesco A. Evangelista}
\email{francesco.evangelista@emory.edu}
\affiliation{Department of Chemistry and Cherry Emerson Center for Scientific Computation, Emory University, Atlanta, Georgia 30322, USA}

\author{Garnet Kin-Lic Chan}
\email{gkc1000@gmail.com}
\affiliation{Division of Chemistry and Chemical Engineering, California Institute of Technology, Pasadena, California 91125, USA}

\author{Gustavo E. Scuseria}
\affiliation{Department of Chemistry, Rice University, Houston, Texas 77005, USA}
\affiliation{Department of Physics and Astronomy, Rice University, Houston, Texas 77005, USA}

\date{\today}

\begin{abstract}
A formal analysis is conducted on the exactness of various forms of unitary coupled cluster (UCC) theory based on particle-hole excitation and de-excitation operators.
Both the conventional single exponential UCC parameterization and a disentangled (factorized) version are considered.
We formulate a differential cluster analysis to determine the UCC amplitudes corresponding to a general quantum state. The exactness of  conventional UCC (ability to represent any state)
is explored numerically and it is formally shown to be determined by the structure of the critical points of the UCC exponential mapping.
A family of \theoryname wave functions are shown to exactly parameterize any state, thus showing how to construct Trotter-error-free parameterizations of UCC for applications in quantum computing.
From these results, we derive an exact \theoryname parameterization that employs an infinite sequence of particle-hole or general one- and two-body substitution operators.
\end{abstract}

\maketitle

\section{Introduction}
Recent developments of  electronic structure quantum algorithms\cite{Whitfield:2011bz,Peruzzo:2014kca,OMalley:2016dc,McClean:2016bs,Shen:2017cc,Xia:2017dl,Xia:2018jn,Hempel:2018ip,Barkoutsos:2018hm,Ryabinkin:2018jw,Lee:2018cy,DallaireDemers:2019iw,Grimsley:2019ed} for noisy intermediate-scale quantum (NISQ) devices\cite{Preskill:2018gt} have renewed interest in the unitary formulation of coupled cluster theory (UCC) and related formalisms.\cite{theory:1977bs,Kutzelnigg:1983wa,Bartlett:1989th,Kutzelnigg:1991vl,Kutzelnigg:2010uv,Taube:2006uy,Harsha:2018dv}
Unitary coupled cluster theory expresses a generic state ($\Psi$) using the exponential ansatz
\begin{equation}
\label{eq:ucc_ansatz}
\ket{\Psi_\mathrm{UCC}} = e^{\hat{\sigma}} \ket{\Phi_0},
\end{equation}
where $\Phi_0$ is a reference Slater determinant and the operator $\hat{\sigma}$ is antihermitian.
Following coupled cluster theory,\cite{Bartlett:2007kv} it is customary to parameterize $\hat{\sigma}$ in terms of a particle-hole excitation operator ($\hat{T}$) that promotes electrons from the hole (occupied) to the particle (unoccupied) orbitals of the reference $\Phi_0$,
\begin{equation}
\label{eq:sigma}
\hat{\sigma}= \hat{T} - \hat{T}^\dagger.
\end{equation}
Although UCC was proposed\cite{Yaris:1965bg,theory:1977bs,Suzuki:1982bq} about a decade after the introduction of coupled cluster theory,\cite{Coester:1958vr,Coester:1960dq,Cizek:1966th,Crawford:2000ub,Shavitt:2009uo} apart from a few exceptions,\cite{Bartlett:1989th,Kutzelnigg:1991vl,White:2002uj,Yanai:2006gi, Yanai:2007ix,Cooper:2010ck,Evangelista:2011eh,Chen:2012bm,Evangelista:2014wu} it has found little application in electronic structure theory  because the corresponding equations cannot be efficiently evaluated without approximation on a classical computer.
However, as suggested by Peruzzo \textit{et al.},\cite{Peruzzo:2014kca} the UCC wave function can be efficiently generated on a quantum computer as a series of quantum gates that implement unitary rotations.
For ease of implementation, it is common to factorize the exponential into a product of smaller unitary rotations via a Trotter approximation (Trotterization) of the exponential.
The Trotterized form of UCC has been employed as an ansatz for the variational quantum eigensolver (VQE),\cite{Peruzzo:2014kca,McClean:2016bs,OMalley:2016dc,Romero:2019hk,Wecker:2015da,Barkoutsos:2018hm}
typically in the singles and doubles approximation (UCCSD).
UCC with $\hat{\sigma}$ made of other than particle-hole excitations has also been investigated.\cite{Grimsley:2019ed}
The definition of UCC in terms of fermionic operators incurs overhead on typical quantum architectures from the fermionic encoding.
Thus hardware-efficient ans\"{a}tze inspired by the UCC hierarchy but which directly substitute the fermionic field operators for spin ladder operators
have also been suggested as a more efficient alternative.\cite{Kandala:2017gh,Ryabinkin:2018jw}
Also, a general scheme to construct states that preserve a number of symmetries (particle number, time-reversal, spin) has recently been presented.\cite{Gard:2019vd}

Interestingly, to the best of our knowledge, the circumstances under which UCC and its variants can exactly parameterize a general fermionic state are not formally established.
When $\hat{\sigma}$ is taken to be a general operator, exactness of the UCC ansatz  is trivial due to the properties of the exponential map.\cite{Shepard:2015hk}
However, the restriction to particle-hole excitations makes the analogous result not obvious.
From the quantum chemistry side, it is difficult to provide a constructive demonstration of the exactness of UCC due to the non-commutativity of the operators entering $\hat{\sigma}$, which prevents the formulation of a cluster analysis.\cite{Laestadius:2019iz}
From the quantum computing side, the UCC ansatz with $N$ parameters is different from a sequence of $N$ elementary unitary gates, and the form of the gates (particle-hole) does not correspond to a standard set of universal gates.

In traditional coupled cluster theory, a comparison of the determinants that enter in $\Psi$ and those generated by the exponential wave function allows to explicitly write equations that express the operator $\hat{T}$ as a function of the coefficients of determinants in $\Psi$. This mapping exists as long as $\braket{\Psi|\Phi_0} \neq 0$ and it is one-to-one, meaning that the operator $\hat{T}$ that represents any given state in the CC form is unique.\cite{Laestadius:2019iz}
The same one-to-one mapping between a state and the operator $\hat{\sigma}$ will not hold for UCC.
For example, in a system containing two electrons in two spatial orbitals of different symmetry $\{\phi_i, \phi_a\}$ and with a reference determinant defined as $\ket{\Phi_0} = \ket{\phi_{i_\alpha} \phi_{i_\beta}}$, the UCC wave function of the same symmetry as $\Phi_0$ may be written as $\exp(\hat{\sigma})\ket{\Phi_0} = \ket{\Phi_0} \cos \theta  + \ket{\Phi^{a_\alpha a_\beta}_{i_\alpha i_\beta}} \sin \theta$, where $\theta$ is the amplitude corresponding to the double excitation $\phi_{i_\alpha} \phi_{i_\beta} \rightarrow \phi_{a_\alpha} \phi_{a_\beta}$.
This wave function can represent any state spanned by the basis $\{\Phi_0, \Phi^{a_\alpha a_\beta}_{i_\alpha i_\beta}\}$, but it has an infinite number of equivalent representations periodic in $2\pi$.

This work has three goals. Firstly, we formulate a cluster analysis of unitary CC based on a differential formalism.
Our approach expresses the UCC operator $\hat{\sigma}$ as an integral along a path that connects the reference to a general state $\Psi$, and we give an algorithm to construct $\hat{\sigma}$ explicitly for a given state.
Under very mild assumptions regarding the nature of the singular points along such paths, this gives a constructive demonstration of the exactness of UCC.
This analysis is supplemented by a series of numerical experiments in which we verify the exactness of UCC on randomly sampled states and perform an analysis of the dimensionality of the set of singular points.
This analysis has similarities with a study of the geometry of quantum computation,\cite{Nielsen:2006fk} with the difference that herein we are focused on the evolution of a state and not an operator.

Our second goal is to study the following \theorypre form of UCC
\begin{equation}
\label{eq:fucc_ansatz}
\ket{\Psi_\mathrm{dUCC}} = \prod_{i} e^{\hat{\sigma}_{\mu_i}} \ket{\Phi_0},
\end{equation}
where the product contains all the UCC terms exactly once and $\hat{\sigma}_{\mu_i}$ denotes a special ordering of the antihermitian operators.
We prove that a family of reorderings exists such that the \theoryname ansatz [Eq.~\eqref{eq:fucc_ansatz}] can exactly represent any state.
Contrary to the view that considers Eq.~\eqref{eq:fucc_ansatz} as a low-order Trotter approximation ($M = 1$) of UCC,\cite{Peruzzo:2014kca,Wecker:2015tb,Barkoutsos:2018hm}
our result shows that the \theoryname should be viewed as a family of alternative \textit{exact} and general parameterizations of fermionic states. 
In the theory of Lie groups, both the conventional unitary transformation of UCC and \theoryname are employed to parameterize elements of a group, and
are referred to as using canonical coordinates of the ``first kind'' and ``second kind,'' respectively.\cite{Wei:1963ki,Wei:1964go}
Thirdly, from the exactness of \theoryname we show that any state may be  represented as a product of one- and two-body particle-hole unitary operators acting on a Slater determinant.
This result is partially related to the analysis of coupled cluster theory with generalized singles and doubles (CCGSD) by Nooijen\cite{Nooijen:2000wv} and Nakatsuji.\cite{Nakatsuji:2000vga}
Our study concludes with an analysis of the CC, UCC, and \theoryname wave functions of a toy model consisting of two electrons in two orbitals.
This model is used to illustrate several interesting features of these wave functions and their significant differences.

\section{Notation}

In this section we will introduce some essential elements of the notation adopted in this work.
Throughout the paper, indices $i,j,\ldots$ ($a,b,\ldots$) label occupied (virtual) orbitals of $\Phi_0$, respectively. General orbital indices are indicated with $p,q,\ldots$.
A generic state $\Psi$ may be written as a full configuration interaction (FCI) expansion as
\begin{equation}
\ket{\Psi} = (c_0 + \sum_{i}^\mathrm{occ} \sum_{a}^\mathrm{vir} c_{i}^{a} \sqop{a}{i} + \frac{1}{4}\sum_{ij}^\mathrm{occ} \sum_{ab}^\mathrm{vir} c_{ij}^{ab} \sqop{ab}{ij} + \ldots) \ket{\Phi_0},
\end{equation}
where a generic second quantized excitation operator is defined as $\sqop{ab\cdots}{ij\cdots} = \cop{a} \cop{b} \cdots \aop{j} \aop{i}$ and $c_{ij\cdots}^{ab\cdots}$ is the corresponding coefficient.
The number of terms in the FCI expansion is indicated with $N_\mathrm{FCI}$.

The operator $\hat{\sigma}$ that enters in the definition of the UCC wave function is written using the compact notation
\begin{equation}
\label{eq:sigma_def}
\hat{\sigma} = \sum_\mu^{\mathrm{exc}} \hat{\sigma}_\mu = \sum_{\mu}^\mathrm{exc} (t_{\mu}\hat{\tau}_\mu - t^*_{\mu}\hat{\tau}^\dagger_\mu)
\end{equation}
where the multi-index $\mu = ((i,j,\ldots),(a,b, \ldots))$ runs over all unique particle-hole excitations (exc), $t_\mu \equiv t_{ij\cdots}^{ab\cdots}$ is a cluster amplitude, and $\hat{\tau}_\mu \equiv \sqop{ab\cdots}{ij\cdots}$ is a shorthand notation for an excitation operator.
For real wave functions we just can restrict our analysis to the case of real amplitudes (orthogonal transformations) and write $\hat{\sigma}$ as
\begin{equation}
\begin{split}
\hat{\sigma}= \sum_{\mu}^\mathrm{exc} t_{\mu}(\hat{\tau}_\mu - \hat{\tau}^\dagger_\mu) = \sum_{\mu}^\mathrm{exc} t_{\mu}\hat{\kappa}_\mu,
\end{split}
\end{equation}
where $\hat{\kappa}_\mu = \hat{\tau}_\mu - \hat{\tau}^\dagger_\mu$ is the antihermitian combination of a pair of excitation/de-excitation operators.
We will work exclusively with real wave functions below, although the results can be easily generalized to the complex case.

\section{Sufficient conditions for the exactness of unitary exponential parameterizations}
\label{sec:sufficient_conditions}
 
In this section we outline sufficient conditions to obtain a representation of an arbitrary state $\Psi$ in the UCC form and  provide an algorithm to perform a UCC cluster analysis to construct the operator $\hat{\sigma}$.
Consider a continuous $s$-dependent path $\Psi(s)$ with  $s \in [0,1]$ that connects the reference $\Phi_0 = \Psi(0)$ (initial state) to a general state $\Psi = \Psi(1)$ (final state).
We require $\Psi(s)$ be normalized and no other restrictions are imposed on the path.
If UCC can represent $\Psi(s)$, then we can write it in exponential form as
\begin{equation}
\label{eq:ucc_path}
\ket{\Psi(s)} = e^{\hat{\sigma}(s)} \ket{\Phi_0}, \quad s \in [0,1],
\end{equation}
where $\hat{\sigma}$ depends on $s$.
Taking the derivative of Eq.~\eqref{eq:ucc_path} with respect to $s$  we can express $\hat{\sigma}(s)$ as a solution of the following differential equation
\begin{equation}
\label{eq:pathucc}
\frac{d}{ds}\ket{\Psi(s)} = \frac{d}{ds}e^{\hat{\sigma}(s)} \ket{\Phi_0}
= \sum_\mu \frac{d t_\mu(s)}{d s} \frac{\partial}{\partial t_\mu(s)} e^{\hat{\sigma}(s)} \ket{\Phi_0},
\end{equation}
with initial condition $\hat{\sigma}(0) = 0$.
Equation~\eqref{eq:pathucc} is an implicit ordinary differential equation for the UCC amplitudes, which we will refer to as the \uccf.
The existence of a solution of Eq.~\eqref{eq:pathucc} for any final state $\Psi$ is a sufficient condition for the UCC parameterization to be exact.

\begin{figure}[h]
   \centering
   \includegraphics[width=3.5in]{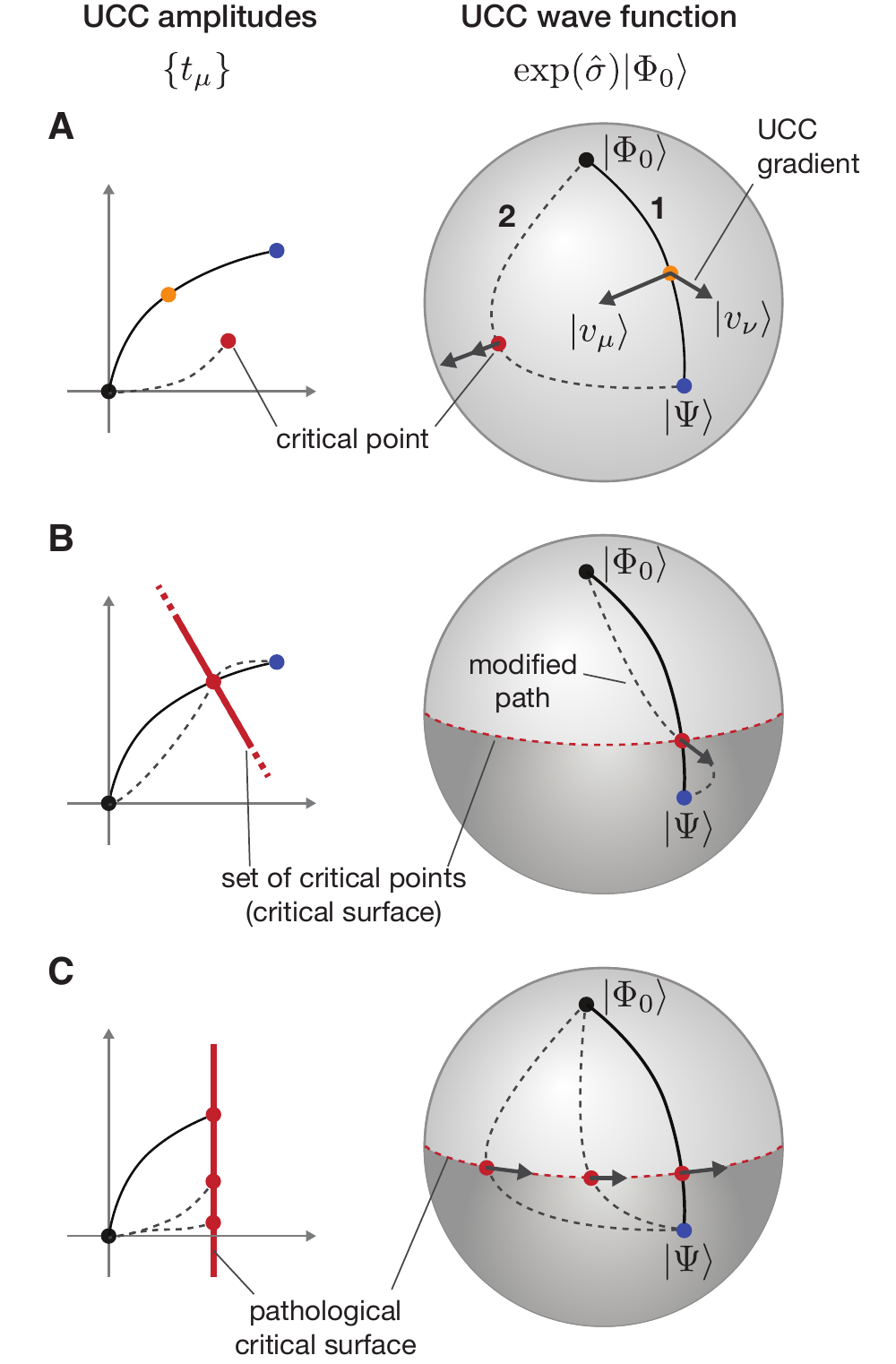}
   \caption{Analysis of various scenarios encountered in the integration of the \uccf [Eq.~\eqref{eq:pathucc}]. Panels on the left visualize the UCC amplitudes while those on the right show the UCC wave function. (A) A path connecting the Slater determinant $\Phi_0$ to the state $\Psi$ may avoid (\textbf{1}) or encounter  (\textbf{2}) a critical point. In the latter case, the UCC gradients $\ket{v_\mu} = \frac{\partial}{\partial t_\mu}\exp(\hat{\sigma})\ket{\Phi_0}$ are linearly dependent. (B) When $\Psi$ is surrounded by a set of critical points (critical surface), a path may still be found that connects $\Phi_0$ to $\Psi$ by making sure that at the critical point the UCC gradients span $\frac{d}{ds} \Psi(s)$. (C) A pathological critical surface cannot be crossed by a path since at each point the UCC gradient is orthogonal to $\frac{d}{ds} \Psi(s)$.}
   \label{fig:ucc_critical_points}
\end{figure}

For a given path $\Psi(s)$, a condition sufficient for the existence of solutions of Eq.~\eqref{eq:pathucc} is that basis of partial derivatives
\begin{equation}
\ket{v_\mu} = \frac{\partial}{\partial t_\mu(s)} e^{\hat{\sigma}(s)} \ket{\Phi_0},
\end{equation}
can represent the gradient $\frac{d}{ds} \Psi(s)$ for all the values of $t_\mu(s)$ taken along the path  (see Fig.~\ref{fig:ucc_critical_points}).
The maximum rank of the basis $\ket{v_\mu}$ is $N_\mathrm{FCI} - 1$, which is equal to the dimensionality of the tangent space of normalized paths $\ket{\Psi(s)}$.
When the basis $\ket{v_\mu}$ is linearly independent (has maximum rank) along the entire path, then the \uccf admits a solution.

The \uccf may fail to have a solution if a path encounters a critical point. Critical points correspond to values of the amplitudes $\{ \tilde{t}_\mu\}$ for which the basis $\ket{v_\mu}$ (or equivalently, the Jacobian of the UCC wave function) has rank smaller than its maximum value.
The critical points of the general exponential map are well characterized in the theory of Lie groups,\cite{RossmannSection:2002te} however, singularities that arise from the particle-hole excitation constraint cannot be addressed by this formalism.
Critical points are potentially problematic, since at such a point the basis of derivatives may not be able to span $\frac{d}{ds} \Psi(s)$.
Since the set of critical points (critical set) corresponds to a subset of the amplitudes that satisfy a constraint, in the worst case scenario it is composed of surfaces of dimension less than $N_\mathrm{FCI} - 1$.
This observation suggests that if a path encounters a critical point, we may use the freedom in choosing the path to find an alternative one that connects to the final state $\Psi$ and avoids critical points.
This case is illustrated in panel A of Fig.~\ref{fig:ucc_critical_points}, where path \textbf{1} avoids critical points while path \textbf{2} encounters a critical point at which the gradients $\ket{v_\mu}$ are linearly dependent.
The existence of isolated critical points is not problematic as one can always find a path that avoids them.

If crossing a critical point is unavoidable (e.g., if the final state is surrounded by a surface of critical points), then one may still be able to modify the path as long as the derivative $\frac{d}{ds} \Psi(s)$ has nonzero overlap with at least one vector $\ket{v_\mu}$.
In this case the path is modified so that at the critical point it is fully spanned by the linearly-independent components of the vectors $\ket{v_\mu}$, e.g., choosing $\frac{d}{ds} \ket{\Psi(s)} = \ket{v_\mu}$.
This second scenario is illustrated in panel B of Fig.~\ref{fig:ucc_critical_points}.
The only case in which the \uccf cannot be integrated occurs if no path can be found that crosses a critical surface. This scenario can arise if all points on a critical surface have gradients $\ket{v_\mu}$ with zero projection on the final state $\Psi$. In this pathological scenario, illustrated in panel C of Fig.~\ref{fig:ucc_critical_points}, once a path reaches the critical set it cannot move in the direction of $\Psi$.
The occurrence of this case seems unlikely in conventional UCC since the exponential mapping from amplitudes to state is highly nonlinear and all amplitudes are coupled.
However, in Sec.~\ref{sec:toymodel} we consider a \theoryname ansatz that is not exact and its analogous path equation displays a pathological critical set.

A precise analytical characterization of the nature and prevalence of critical points in UCC appears to be challenging.
We can also examine the existence of critical points and the issue of integrability of Eq.~\eqref{eq:pathucc} with a series of numerical experiments.
After projection on the left with a set of determinants $\Phi_\mu$, Eq.~\eqref{eq:pathucc} may be expressed as
\begin{equation}
\label{eq:flowucc_linear_system}
\mathbf{c}(s) = \mathbf{A}(s) \dot{\mathbf{t}}(s),
\end{equation}
where the component of the vectors $\mathbf{c}(s)$, $\dot{\mathbf{t}}(s)$ are  $c_\mu(s) = \braket{\Phi_\mu|\frac{d}{ds}\Psi(s)}$ and $\dot{t}_\mu(s) = \frac{d t_\mu(s)}{ds}$, while the matrix $\mathbf{A}(s)$ defined as $A_{\mu\nu}(s) = \braket{\Phi_\mu|v_\nu(s)} = \frac{\partial}{\partial t_\nu(s)}\bra{\Phi_\mu} e^{\hat{\sigma}(s)} \ket{\Phi_0}$ is the Jacobian of the vector valued function  $f_{\mu}(\{ t_\nu\}) = \bra{\Phi_\mu} e^{\hat{\sigma}} \ket{\Phi_0}$.
Note that $\mathbf{A}(s)$ is a rectangular matrix of dimension $N_\mathrm{FCI}\times(N_\mathrm{FCI} - 1)$, so that Eq.~\eqref{eq:flowucc_linear_system} appears to be an overdetermined system of equations.
In practice, we must account for the additional condition $\bra{\Psi(s)}\frac{d}{ds}\ket{\Psi(s)}= 0$, which we consider by solving a constrained linear system, whose solution is given by
\begin{equation}
\label{eq:flow_step_solution}
\dot{\mathbf{t}}(s) = [\mathbf{A}^T(s)\mathbf{A}(s)]^{-1} \mathbf{A}^T(s) \mathbf{c}(s),
\end{equation}
and integrate this equation numerically.

In a first set of experiments, we sample various states $\Psi$ and numerically integrate the \uccf.
Our computations consider six electrons distributed in six orbitals (12 spin orbitals) imposing the constraint $M_S = 0$, which results in a FCI space containing 400 determinants.
The \uccf was solved using as a final state $\Psi$ five types of solutions: a) 100 uniformly distributed random states, b) 100 random states with a small number of determinants that have large weights, c) 10 random seniority zero states (i.e., the states are composed only of determinants with electrons paired in orbitals), d) a seniority zero state with all determinant coefficients equal, and e) all possible
single determinant states in the orbital basis.
Integration of the \uccf for all these cases proceeded without numerical issues sampling ten points along the path for a total of more than 7000 states. At each point along all paths, the Jacobian was found to be of full rank (399), implying that no critical point was visited.
In all cases examined we find that the cluster amplitudes  lie in the range $t_\mu \in [-\pi,\pi]$, with the extremal values $t_\mu = \pm \pi$ encountered only when the final state is $\ket{\Psi} = -\ket{\Phi_0}$.
When the final state is an excited determinant, $\Psi = \Phi_\mu$ ($\mu > 0$) then the corresponding amplitude takes the value $|t_\mu| = \pi/2$.

In a second set of numerical experiments, we identified and analyzed the nature of critical points of a system of four electrons in four orbitals, where $\Phi_0$ is taken to be a fixed determinant with all electrons paired.
We identified 100 critical points by numerical minimization of the smallest singular value of $\mathbf{A}(s)$ ($\sigma_\mathrm{min}$).
To characterize the neighborhood of these points, we expand $\sigma_\mathrm{min}$ around each  critical point $\{ \tilde{t}_\mu\}$ in a Taylor series
\begin{equation}
\sigma_\mathrm{min}(\{t_\mu\}) \approx \sigma_\mathrm{min}(\{\tilde{t}_\mu\}) + \frac{1}{2} \sum_{\alpha,\beta} \frac{\partial^2 \sigma_\mathrm{min}(\{\tilde{t}_\mu\})}{\partial t_\alpha \partial t_\beta}  (t_\alpha - \tilde{t}_\alpha) (t_\beta - \tilde{t}_\beta),
\end{equation}
where we have taken into account that gradient terms are zero at a critical point.
If all the eigenvalues of the Hessian $\frac{\partial^2 \sigma_\mathrm{min}(\{\tilde{t}_\mu\})}{\partial t_\alpha \partial t_\beta}$ are positive, then any small change in the amplitudes will make $\sigma_\mathrm{min}(\{t_\mu\}) > 0$ and, therefore, $\mathbf{A}(s)$ of full rank.
Instead, if $n$ eigenvalues are zero, then there are $n$ directions in amplitude space that keep $\sigma_\mathrm{min}(\{t_\mu\}) = 0$, and consequently $\mathbf{A}(s)$ rank deficient.
In all cases, we found the eigenvalues of the Hessian to be positive, which implies that all sampled critical points have dimension zero.

In summary, under the assumption that the set of paths connecting the reference and desired wave function does not display a pathological critical set, we provide an explicit algorithm to construct the exact UCC representation.
We explore this construction in our numerical experiments, which show that UCC solutions can be found for a variety of states and we find here that the critical set consists of isolated points of dimension zero.
These results also find confirmation in our study of the UCC wave function for a toy model reported in Sec.~\ref{sec:toymodel}.

\section{Exactness of a \theorypre form of particle-hole unitary coupled cluster theory}
\label{sec:ducc}

\begin{figure*}[htbp]
   \centering
   \includegraphics[width = 6in]{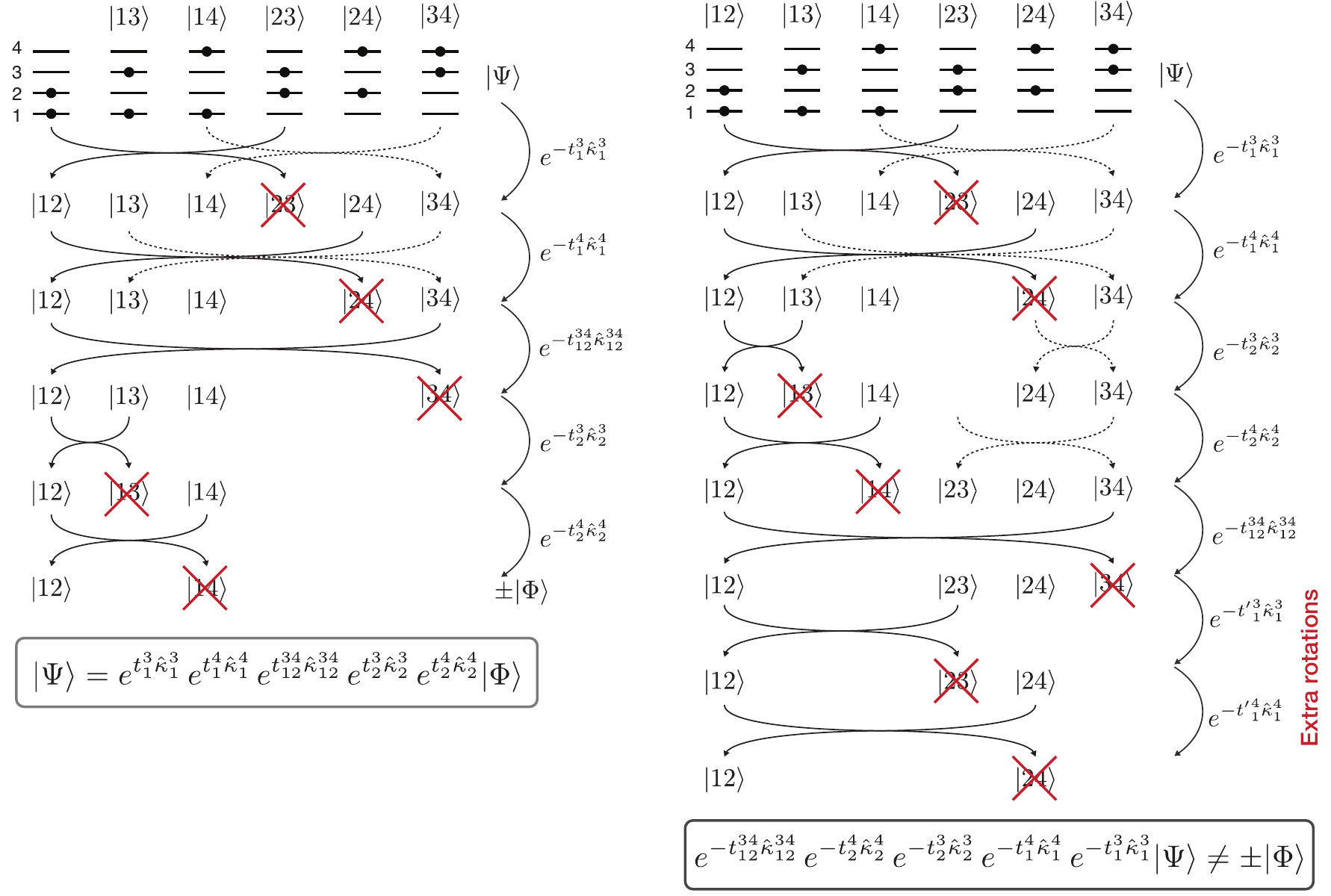}
   \caption{Sequence of elementary unitary transformations that rotate a general  two electrons in four spin orbitals state $\Psi$ to the reference determinant $\Phi_0 = |12\rangle$.
   At each step, a rotation is applied to eliminate a determinant from the current state. Determinants eliminated after a rotation are marked by a red cross.
   Arrows that connect pairs of determinants indicate the effect of each single rotation. Rotations that do not project out determinants are indicated with dashed lines.}
   \label{fig:fucc}
\end{figure*}

In this section we prove that the \theorypre form of the UCC ansatz [Eq.~\eqref{eq:fucc_ansatz}] can exactly represent any state, provided that an appropriate ordering $\{ \mu_i , i = 1, \ldots, N_\mathrm{FCI}\}$ of the sequence of unitary transformations is chosen.
We explicitly show how to construct such sequences and provide an algorithm to perform a \theoryname cluster analysis.
Note that although we use the same symbol to indicate cluster amplitudes entering into the UCC and dUCC wave functions, these have in principle distinct numerical values.

To prove that any state may be written in the \theoryname form, we equate this ansatz to a general wave function and apply the inverse of the product of exponential operators to both sides:
\begin{equation}
\label{eq:ducc_proof_statement}
\left(\prod_{i}^{N_\mathrm{exc}} e^{t_{\mu_i} \hat{\kappa}_{\mu_i}}\right)^{-1} \ket{\Psi} = \cdots e^{-t_{\mu_2} \hat{\kappa}_{\mu_2}}   e^{-t_{\mu_1} \hat{\kappa}_{\mu_1}} \ket{\Psi} = \ket{\Phi_0}.
\end{equation}
Equation~\eqref{eq:ducc_proof_statement} connects the exactness of the \theoryname ansatz to the existence of a product of unitary operations that rotates the components of $\Psi$ into the determinant $\Phi_0$.

A systematic transformation that realizes Eq.~\eqref{eq:ducc_proof_statement} may be built by a sequence of rotations that progressively eliminates all determinants labeled by one occupied orbital at a time.
Consider an occupied index $i$. We start by removing determinants that are singly excited with respect to the reference (singles) by unitary (singles) transformations of the form
\begin{equation}
\exp(- t_{i}^{a} \hat{\kappa}_{i}^{a}).
\end{equation}
For each term, the amplitude $t_{i}^{a}$ is determined by requiring that the state produced after its application
\begin{equation}
\ket{\Psi'} = \exp(- t_{i}^{a} \hat{\kappa}_{i}^{a}) \ket{\Psi}
\end{equation}
does not contain the corresponding excited determinant $\Phi_{i}^{a}$, that is, $\braket{\Phi_{i}^{a} | \Psi'} = 0.$
The operator $\exp(- t_{i}^{a} \hat{\kappa}_{i}^{a})$ rotates the reference and  excited determinants according to
\begin{equation}
\label{eq:single_ducc_rotations}
\begin{split}
\exp(- t_{i}^{a} \hat{\kappa}_{i}^{a}) \ket{\Phi_0} &= \ket{\Phi_0} \cos(t_{i}^{a}) - \ket{\Phi_{i}^{a}}\sin(t_{i}^{a}), \\
\exp(- t_{i}^{a} \hat{\kappa}_{i}^{a}) \ket{\Phi_{i}^{a}} &= \ket{\Phi_{i}^{a}} \cos(t_{i}^{a}) + \ket{\Phi_0}\sin(t_{i}^{a}), \\
\exp(- t_{i}^{a} \hat{\kappa}_{i}^{a}) \ket{\Phi_{ij}^{ab}} &= \ket{\Phi_{ij}^{ab}} \cos(t_{i}^{a}) + \ket{\Phi_{j}^{b}}\sin(t_{i}^{a}), \\
&\;\;\vdots
\end{split}
\end{equation}
Therefore, the resulting rotated state is modified to
\begin{equation}
\begin{split}
\ket{\Psi'} = & [c_0  \cos(t_{i}^{a}) + c_{i}^{a} \sin(t_{i}^{a}) ]\ket{\Phi_0}
+ \ldots \\
& + [-c_0  \sin(t_{i}^{a}) + c_{i}^{a} \cos(t_{i}^{a}) ]\ket{\Phi_{i}^{a}} + \ldots,
\end{split}
\end{equation}
and the determinant $\Phi_{i}^{a}$ may be eliminated by selecting
\begin{equation}
\label{eq:single_ducc_amplitude}
t_{i}^{a} = \arctan\left(\frac{c_{i}^{a}}{ c_0 }\right).
\end{equation}
If $c_0 = 0$, both $t_{i}^{a} = \pm \pi/2$ solutions are acceptable.
After eliminating $\Phi_{i}^{a}$ from $\Psi$, one may proceed in a similar way with all other singly excited determinants of the form $\Phi_{i}^{b}$, $\Phi_{i}^{c}$, $\ldots$, with $b \neq a, c \neq a, \ldots$.
It is important to note that each successive singles transformation does not reintroduce previously eliminated singles. For example, once $\Phi_{i}^{a}$ is eliminated from $\Psi$ it cannot be generated by $\hat{\kappa}_{i}^{b}$ acting on any of the determinants contained in $\Psi'$.

After all singles labeled by orbital $i$ are removed, one proceeds to suppress double excitations of the form $\Phi_{ij}^{ab}$ via the unitary rotations $\exp(- t_{ij}^{ab} \hat{\kappa}_{ij}^{ab})$.
A set of equations similar to Eqs.~\eqref{eq:single_ducc_rotations}--\eqref{eq:single_ducc_amplitude} may be written to determine the value of $t_{ij}^{ab}$ that removes $\Phi_{ij}^{ab}$.
Like in the case of singles, an important aspect to point out is that this rotation does not reintroduce any of the single excitations previously removed because no triply excited determinant can be de-excited by an operator of the form $\hat{\kappa}_{ij}^{ab}$ to give $\Phi_{i}^{a}$.
This procedure can be carried forward until doubles and all higher excitations in which one occupied orbital is labeled by the index $i$ are rotated out.

Once this sequence of transformations is applied, all excited determinants in the FCI expansion labeled by the index $i$ are removed and the coefficients of the remaining determinants will be different from those of the initial state $\Psi$.
This procedure may be repeated choosing the next occupied orbital index $j \neq i$. As determinants labeled are $j$ removed by sequences of rotations, determinants labeled by $i$ are not reintroduced in the transformation since they were completely removed in the previous step. Continued iteration over the remaining occupied orbital indices allows to remove all excited determinants from $\Psi$ leading in the end to the state $\pm\Phi_0$.
Given the freedom in the choice of the order in which excited determinants are removed, there are many inequivalent exact \theoryname ans\"{a}tze that
can be constructed by this procedure.

The procedure described here to reduce a general state to the \theoryname form is illustrated in Fig.~\ref{fig:fucc} for a system of two electrons in four spin orbitals $\{ \psi_p, p = 1, \ldots, 4 \}$.
In this case a generic state may be written as $\ket{\Psi} = \sum_{p < q} c_{pq} \ket{pq}$, where $\ket{pq} \equiv \ket{\psi_p \psi_q}$ denotes a determinant.
In Fig.~\ref{fig:fucc} A we show the sequence of inverse unitaries corresponding to the exact \theoryname wave function
\begin{equation}
e^{t_{1}^{3} \hat{\kappa}_{1}^{3}} \,
e^{t_{1}^{4} \hat{\kappa}_{1}^{4}} \,
e^{t_{12}^{34} \hat{\kappa}_{12}^{34}} \,
e^{t_{2}^{3} \hat{\kappa}_{2}^{3}} \,
e^{t_{2}^{4} \hat{\kappa}_{2}^{4}} |\Phi_0\rangle.
\end{equation}
Several determinants may be affected by an operator during each step of the inverse transformation. However, the sequence of operations is guaranteed to eliminate one determinant without reintroducing previously removed determinants.
In Fig.~\ref{fig:fucc} B we show a different \theoryname wave function that cannot exactly represent a general state, namely
\begin{equation}
e^{t_{12}^{34} \hat{\kappa}_{12}^{34}} \,
e^{t_{1}^{3} \hat{\kappa}_{1}^{3}} \,
e^{t_{1}^{4} \hat{\kappa}_{1}^{4}} \,
e^{t_{2}^{3} \hat{\kappa}_{2}^{3}} \,
e^{t_{2}^{4} \hat{\kappa}_{2}^{4}} |\Phi_0\rangle.
\end{equation}
In this case, if we perform a cluster analysis, the application of two single rotation operators [$\exp(-t_{2}^{3} \hat{\kappa}_{2}^{3})$ and $\exp(-t_{2}^{4} \hat{\kappa}_{2}^{4})$] reintroduces determinants $\ket{23}$ and $\ket{24}$, which were removed in the preceding two steps.
Application of two extra unitary operations [$\exp(-{t'}_{1}^{3} \hat{\kappa}_{1}^{3})$ and $\exp(-{t'}_{1}^{4} \hat{\kappa}_{1}^{4})$] eliminates all excited determinants.
However, an ansatz including these extra operators is not minimal as it requires more disentangled particle-hole unitaries (7) than required by the exact ansatz (5).

\section{Exact product of one- and two-body particle-hole unitary ansatz}
\label{sec:nooijen}

\begin{figure}[tbp]
   \centering
   \includegraphics[width=3.25in]{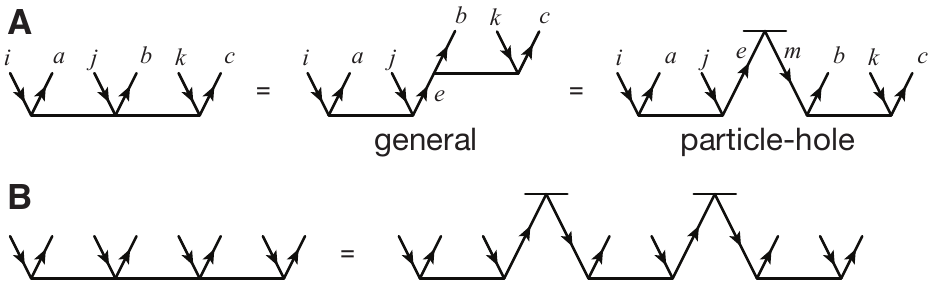}
   \caption{Diagrammatic depiction of the decomposition of the particle-hole three- (A) and four-body (B) operators $\sqop{abc}{ijk}$ and $\sqop{abcd}{ijkl}$ using one- and two-body operators.}
   \label{fig:commutator}
\end{figure}

Using the exactness result from Sec.~\ref{sec:ducc}, we show that any state may be approximated with arbitrary precision by an infinite product of only \textit{particle-hole} one- and two-body unitary operators. 
We start by noting that an arbitrary $n$-body term of the \theoryname ansatz [Eq.~\eqref{eq:fucc_ansatz}]
\begin{equation}
\label{eq:single_unitary}
t^{a_1 \ldots a_n}_{i_1 \ldots i_n} \sqrot{a_1 \ldots a_n}{i_1 \ldots i_n},
\end{equation}
may be decomposed into a sequence of commutators of one- and two-body particle-hole antihermitian operators.
For example, the three-body operator $\sqrot{abc}{ijk}$ may be written as a single commutator of two \textit{general} two-body operators $\sqrot{abc}{ijk} = -[\sqrot{a e}{ij},\sqrot{bc}{ek}]$ ($e \notin \{a, b,c\}$), or as a doubly nested commutator involving only \textit{particle-hole} operators $\sqrot{e}{m}$ and $\sqrot{bc}{mk}$
\begin{equation}
\sqrot{abc}{ijk}
= -[\sqrot{a e}{ij}, [\sqrot{e}{m},\sqrot{bc}{mk}]], \quad e \notin \{a,b,c\}.
\end{equation}
These two decompositions of the excitation component of $\sqrot{abc}{ijk}$ are illustrated using standard diagrammatic notation in Fig.~\ref{fig:commutator}.
In the same figure, we also show that this factorization generalizes to higher-order excitation operators, for example, $\sqrot{abcd}{ijkl}$.
An $n$-body operator $\sqrot{a_1 \ldots a_{n}}{i_1 \ldots i_{n}}$ may be recursively broken down as a series of nested commutators using the relation
\begin{equation}
\sqrot{a_1 \ldots a_{n}}{i_1 \ldots i_{n}}
= - [\sqrot{a_1 \ldots e}{i_1 \ldots i_{n-1}},[\sqrot{e}{m},\sqrot{a_{n-1}a_{n}}{m \, i_{n}}], \quad e \notin \{a_1,\ldots,a_n\}.
\end{equation}
When there are more unoccupied orbitals than occupied ones, this recursive decomposition can be applied up to the highest excitation level because it is always possible to choose an index $e \notin \{a_1,\ldots,a_n\}$; if the number of unoccupied and occupied orbitals is equal, one can simply add a fictitious orbital so that the above construction can be performed.

Once a generic operator $\sqrot{a_1 \ldots a_n}{i_1 \ldots i_n}$ is expressed as a series of nested commutators, its exponential may be obtained via repeated application of the following operator identity\cite{Deutsch:1995gi}
\begin{equation}
\label{eq:trotter_commutator}
e^{[A,B]} = \lim_{M \rightarrow \infty} \left(e^{A/\sqrt{M }} e^{B/\sqrt{M }} e^{-A/\sqrt{M }} e^{-B/\sqrt{M }}\right)^M.
\end{equation}
When each exponential term of the \theoryname ansatz is decomposed in this way, it leads to an infinite series of particle-hole one- and two-body antihermitian operators [$\hat{\kappa}^{(1,2)}_{\mu_i} \in \{ \hat{\kappa}_{i}^{a}, \hat{\kappa}_{ij}^{ab} \}$]
\begin{equation}
\label{eq:sequcc_nooijen}
\ket{\Psi_\mathrm{dUCCSD(\infty)}} = 
\prod_{i}^\infty e^{t_{\mu_i} \hat{\kappa}^{(1,2)}_{\mu_i}} \ket{\Phi_0},
\end{equation}
where the indices $\mu_i$ label a specific sequence of one- and two-body operators.
Finite approximations may be derived by expanding Eq.~\eqref{eq:trotter_commutator} up to a given order $M$.
In an analogous way, it is easy to show that the \theoryname wave function may also be decomposed as an infinite product of general one- and two-body unitary operators, $\prod_{i}^\infty e^{t_{\mu_i} \hat{\kappa}^{(1,2)}_{\mu_i}} \ket{\Phi_0}$, with the operators selected from the set $\hat{\kappa}^{(1,2)}_{\mu_i} \in \{ \hat{\kappa}_{p}^{q}, \hat{\kappa}_{pq}^{rs} \}$.

Compared to CCGSD, which despite being numerically very accurate\cite{vanVoorhis:2001us,Piecuch:2003hl,Fan:2006fz,Fan:2006jy} was shown to be an inexact wave function parameterization,\cite{Davidson:2003ch,Ronen:2003ft,Mazziotti:2004bu,Mukherjee:2004jp,Kutzelnigg:2005el}
the \theoryname with singles and doubles can be made arbitrarily accurate by increasing the number of terms in the parameterization.

\section{Example:Two electrons in four spin orbitals}
\label{sec:toymodel}
In this section we analyze the features of UCC and the \theoryname wave function using a simple toy model consisting of two fermions in two spatial orbitals $\phi_i$ and $\phi_a$. For convenience we indicate spin orbitals with the label $i$ and $a$ and use a bar to indicate beta spin functions.
The Hilbert space for this toy model is spanned by the determinants $\{ |i \bar{i} \rangle , |a \bar{i} \rangle,  |i \bar{a} \rangle, |a \bar{a} \rangle\}$ and a general state will be written as
\begin{equation}
\label{eq:toyproblem_fci}
|\Psi\rangle =  c_{1} |i \bar{i} \rangle + c_{2} |a \bar{i} \rangle + c_{3} |i \bar{a} \rangle + c_{4} |a \bar{a} \rangle.
\end{equation}

\begin{figure*}[htbp]
   \centering
   \includegraphics{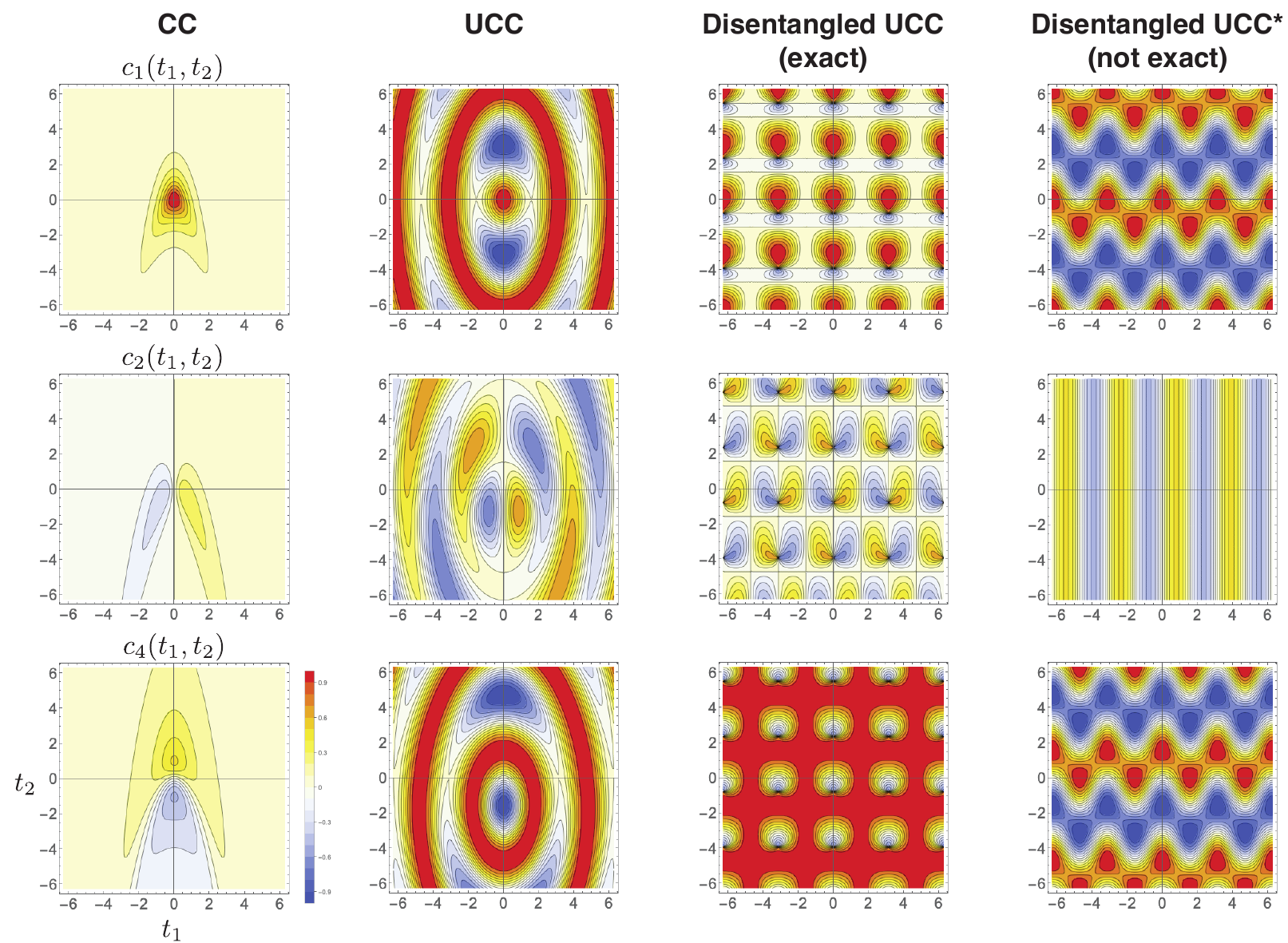}
   \caption{Two electrons in two spatial orbitals toy problem. Contour plots of the FCI wave function coefficients [see Eq.~\eqref{eq:toyproblem_fci}] as a function of the cluster amplitudes $t_1$ and $t_2$ for the coupled cluster (CC) [Eq.~\eqref{eq:toyproblem_cc}], unitary CC (UCC) [Eq.~\eqref{eq:toyproblem_ucc}], and two versions of the \theoryname ansatz [Eqs.~\eqref{eq:toyproblem_ducc_exact} and \eqref{eq:toyproblem_ducc_inexact}]. Contour values range from +1 (in red) to $-1$ (in blue). The coefficients of determinants $\ket{\bar{i}a}$ and $\ket{i\bar{a}}$ are constrained to be equal ($c_2 = c_3$). For the disentangled UCC wave function we show only one of the solutions compatible with this constraint.}
   \label{fig:toyproblem}
\end{figure*}

\subsection{Traditional CC}

Taking the reference state to be $\ket{\Phi_0} = |i \bar{i} \rangle$, we write the cluster operator in terms of cluster amplitudes for singles ($t_1, t'_1$) and doubles ($t_2$) as
\begin{equation}
\hat{T} = t_1 \hat{a}_i^a  + t'_1 \hat{a}_{\bar{i}}^{\bar{a}}
+ t_2 \, \hat{a}_{i\bar{i}}^{a\bar{a}}.
\end{equation}
The coupled cluster wave function is a simple polynomial in $t_1$, $t'_1$, and $t_2$ and in intermediate normalization it reads
\begin{equation}
\label{eq:toyproblem_cc}
c_1 = 1, \quad
c_2 = t_1, \quad
c_3 = t_1', \quad
c_4 = t_2 + t_1 t'_1.
\end{equation}
This system of equations is easily inverted to obtain expressions for $t_1$ and $t_2$ in terms of the FCI coefficients, giving $t_1 = c_2$, $t'_1 = c_3$, and $t_2 = c_4 - c_2 c_3$.
In Fig.~\ref{fig:toyproblem} we plot the determinant coefficients ($c_i$) as a function of the CC amplitudes. For convenience, we impose the constraint $c_2 = c_3$ and normalize the wave function to one.

The exactness of the traditional CC ansatz may also be analyzed from the point of view of a \flow formalism. Writing the CC path as, $\ket{\Psi(s)} = \exp\hat{T}(s) \ket{\Phi_0}$, where the amplitudes are functions of $s$, we can write a corresponding path equation
\begin{equation}
\label{eq:pathcc}
\frac{d}{ds}\ket{\Psi(s)} = \frac{d}{ds}e^{\hat{T}(s)} \ket{\Phi_0}
= \sum_\mu \frac{d t_\mu(s)}{d s} e^{\hat{T}(s)} \ket{\Phi_\mu},
\end{equation}
where we have used the fact that $\hat{a}_\mu$ and $\hat{T}(s)$ commute.
In the CC formalism, the corresponding gradient vectors are given by $\ket{v_\mu} = e^{\hat{T}(s)} \ket{\Phi_\mu}$, and a solution to the CC path equation exists if this basis can span the path derivative $\frac{d}{ds}\ket{\Psi(s)}$. One way to characterize the singular points of this basis is to evaluate the metric matrix $(\mathbf{M})_{\mu\nu} = \braket{v_\mu | v_\nu}$, which may be related to the Jacobian matrix $\mathbf{A}$ via $\mathbf{M} = \mathbf{A}^T \mathbf{A}$.
When the gradient vectors are linearly dependent, $\mathbf{M}$ is singular and its determinant is null.
For the toy model, the CC gradient vectors $\ket{v_\mu}$ and metric are given by
\begin{align}
\ket{v_1} \rightarrow 
\begin{pmatrix}
0 \\ 1 \\ 0 \\ t'_1
\end{pmatrix},
\quad
\ket{v'_{1}} \rightarrow 
\begin{pmatrix}
0 \\ 0 \\ 1 \\ t_1
\end{pmatrix},
\quad
\ket{v_2} \rightarrow 
\begin{pmatrix}
0 \\ 0 \\ 0 \\ 1
\end{pmatrix},
\\
\mathbf{M} = 
\begin{pmatrix}
1 + (t'_1)^2 & t_1 t'_1 & t'_1 \\
t_1 t'_1 & 1 + (t_1)^2 & t_1 \\
t'_1 & t_1 & 1
\end{pmatrix}.
\end{align}
Since in this case, independently of the value of the amplitudes, $\det \mathbf{M} = 1$, there are no singular points in the CC exponential mapping leaving considerable freedom in the choice of the path. However, the gradient vectors have no overlap with the reference ($\braket{v_\mu | \Phi_0} = 0$ for all $\mu$), so the CC path is integrable provided that the path satisfies intermediate normalization, i.e.,  $\braket{\Psi(s) | \Phi_0} = 1$ for $s \in [0,1]$.

\subsection{Unitary CC}
\begin{figure}[bp]
   \centering
   \includegraphics[width=3.5in]{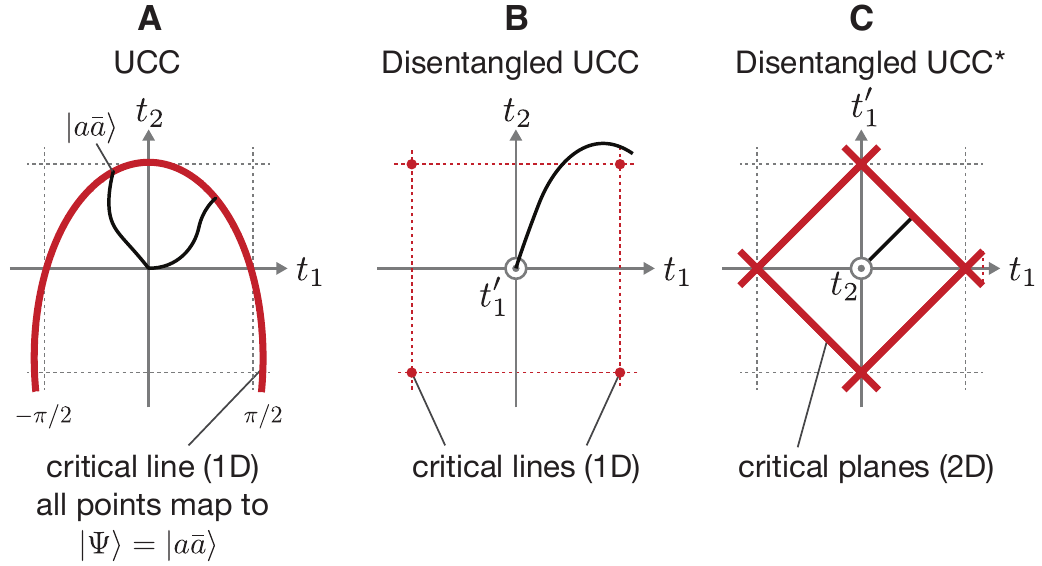}
   \caption{Two electrons in two spatial orbitals toy problem. Set of critical points for the UCC wave function (A), a \theoryname wave function that can represent any state (B), and a \theoryname wave function that cannot represent all states (C). In (A) the first critical line corresponds to the state $\ket{a\bar{a}}$ and paths can cross over it. In (B), due to the low dimensionality of the critical sets, one can always find a path that avoids them. In (C), when trajectories that move towards the state $\tilde{\Psi}$ [Eq.~\eqref{eq:entangled}] approach the critical plane, the gradient of the wave function becomes orthogonal to $\tilde{\Psi}$ and this boundary cannot be crossed.}
   \label{fig:critical}
\end{figure}

For the toy model, if we impose the constraint $c_2 = c_3$ it is possible to derive closed form expressions for the determinant coefficients in the UCC wave function:
\begin{equation}
\label{eq:toyproblem_ucc}
\begin{split}
c_1 &= \frac{2 t_1^2 + (2 t_1^2 + 2 t_2^2) \, \cos(\sqrt{4 t_1^2 + t_2^2})}{4 t_1^2 + t_2^2}, \\
c_2 &= \frac{t_1 \left[t_2 
\left(\cos\sqrt{4 t_1^2 + t_2^2}-1\right) + \sqrt{4 t_1^2 + t_2^2} \sin\sqrt{4 t_1^2 + t_2^2} \right]}{4t_1^2 + t_2^2}, \\
c_3 &= c_2, \\
c_4 &= \frac{2 t_1^2 \left[1 - \cos\sqrt{4 t_1^2 + t_2^2} \right]+ t_2 \sqrt{4 t_1^2 + t_2^2} \sin\sqrt{4 t_1^2 + t_2^2}}{4t_1^2 + t_2^2}.
\end{split}
\end{equation}
As shown in Fig.~\ref{fig:toyproblem}, this wave function has a significantly more complex behavior than that of CC, including multiple representations of a given state.
For this UCC solution, it is possible to write the determinant of the corresponding metric matrix as
\begin{equation}
\det\mathbf{M} = -8 \, \sin\left(\frac{r}{2}\right)^2 \frac{ \
\frac{1}{2}  \cos^2 \theta (1 + \cos r) +\sin^2\theta - \sin r \sin \theta }{r^2}
\end{equation}
where we expressed $t_1$ and $t_2$ in term of polar coordinates ($r$,$\theta$) as
\begin{equation}
t_1 = \frac{1}{2} r \cos \theta, \quad t_2 = r \sin \theta.
\end{equation}
The set of critical values $(t_1,t_2)$ in the domain $|t_\mu| \leq \pi$ ($\mu = 1,2$) is shown in Fig.~\ref{fig:critical} A.
One of the special features of this critical set is that all points that belong to it consist of equivalent representations of the state, $\ket{\Psi} = |a \bar{a} \rangle$.
Thus, any path $\Psi(s)$ that avoids the state $|a \bar{a} \rangle$ is integrable. Moreover, the state $|a \bar{a} \rangle$ can be represented trivially by the UCC ansatz. So UCC can represent any state in the space of stated defined by this toy model.

\subsection{Disentangled UCC}
We begin by considering the following \theoryname wave function, which can exactly represent any state:
\begin{equation}
\label{eq:cas22_ducc_exact}
\ket{\Psi_\mathrm{dUCC}} = 
 e^{t_1' (\hat{a}_{\bar{i}}^{\bar{a}}-\hat{a}^{\bar{i}}_{\bar{a}})} e^{t_2 \, (\hat{a}_{i\bar{i}}^{a\bar{a}} - \hat{a}^{i\bar{i}}_{a\bar{a}})} e^{t_1 (\hat{a}_i^a - \hat{a}^i_a)} \ket{\Phi_0},
\end{equation}
to which corresponds the following determinant coefficients
\begin{equation}
\label{eq:toyproblem_ducc_exact}
\begin{split}
c_1 &= \cos(t_1)\cos(t_1') \cos(t_2), \\
c_2 &= \sin(t_1) \cos(t_1')  - \cos(t_1) \sin(t_1') \sin(t_2), \\
c_3 &= \cos(t_1) \sin(t_1') \cos(t_2), \\
c_4 &= \sin(t_1)\sin(t_1') + \cos(t_1)\cos(t_1') \sin(t_2).
\end{split}
\end{equation}
In writing this ansatz we keep both spin components of the singles amplitudes ($t_1$ and $t_1'$, respectively) since spin conservation is not trivially enforced by the condition $t_1 = t_1'$.
In this case it is possible to invert the equations and express $t_1$, $t_1'$, and $t_2$ as a function of the FCI coefficients as
\begin{equation}
\begin{split}
t_1' & = \arctan \frac{c_3}{c_1},\\
t_1 & = \arcsin[ c_2 \cos(t_1')  + c_4 \sin(t_1')],\\
t_2 & = \arccos\frac{c_1}{\cos(t_1)\cos(t_1')}.
\end{split}
\end{equation}
The exactness of this \theoryname ansatz can also be discussed from the point of view of the \flow equation.
For the ansatz in Eq.~\eqref{eq:cas22_ducc_exact} one finds that the determinant of the metric matrix is given by
\begin{equation}
\det \mathbf{M} = \cos^4(t_1) \cos^2(t_2),
\end{equation}
which implies that critical points correspond to values of $t_1 =\pi /2  + \pi k$ and  $t_2 =\pi /2  + \pi l$, with $k,l \in \mathbb{Z}$.
Critical points of the \flow equation lie on the intersection of these two planes (see Fig.~\ref{fig:critical} B) and correspond to a lattice  of lines (a set of dimension one) with coordinates $(t_1,t'_1,t_2) = (\pi /2  + \pi k, t ,\pi /2  + \pi l)$ where $t \in \mathbb{R}$. Therefore, it is always possible to find a path that avoids the singular points.

A feature of the \theoryname is that the intermediate states generated by the sequence of unitary operators may introduce determinants that are not in the final state.
For example, consider the following entangled state
\begin{equation}
\label{eq:entangled}
\ket{\tilde{\Psi}} = \frac{|a \bar{i} \rangle +  |i \bar{a} \rangle }{\sqrt{2}},
\end{equation}
which corresponds to coefficient values $c_2 = c_3 = 1/\sqrt{2}$ and is  represented in the \theoryname form by the following set of parameters $t_1 = 0$, $t_2 = - \pi / 4$, and $t_1' = \pi/2$.
The sequence of operations that generate this state is:
\begin{equation}
\ket{i \bar{i}}
\xrightarrow{\exp(-\frac{\pi}{4} \, \hat{\kappa}_{i\bar{i}}^{a\bar{a}})}
\frac{1}{\sqrt{2}} (\ket{i \bar{i}} - \ket{a \bar{a}})
\xrightarrow{\exp(\frac{\pi}{2} \, \hat{\kappa}_{\bar{i}}^{\bar{a}})}
\frac{1}{\sqrt{2}} (\ket{i \bar{a}} + \ket{a \bar{i}}),
\end{equation}
and we see that the middle step introduces a doubly excited determinant that does not contribute to the final wave function.

Next, we study a \theoryname ansatz that cannot represent a general state.
Consider the following state, where singles are applied before the doubles
\begin{equation}
\label{eq:cas22_ducc_inexact}
\ket{\Psi_\mathrm{dUCC^*}} = 
e^{t_2 \, (\hat{a}_{i\bar{i}}^{a\bar{a}} - \hat{a}^{i\bar{i}}_{a\bar{a}})} e^{t'_1(\hat{a}_{\bar{i}}^{\bar{a}}-\hat{a}^{\bar{i}}_{\bar{a}})} e^{t_1 (\hat{a}_i^a - \hat{a}^i_a)} \ket{\Phi_0}.
\end{equation}
The corresponding determinant coefficients are given by
\begin{equation}
\begin{split}
\label{eq:toyproblem_ducc_inexact}
c_1 &= \cos(t_1) \cos(t'_1) \cos(t_2) - \sin(t_1) \sin(t'_1) \sin(t_2), \\
c_2 &= \sin(t_1) \cos(t'_1) , \\
c_3 &= \cos(t_1) \sin(t'_1) , \\
c_4 &= \sin(t_1) \sin(t'_1) \cos(t_2) + \cos(t_1) \cos(t'_1) \sin(t_2).
\end{split}
\end{equation}
This second parameterization cannot be exact in general, which we illustrate by showing that it cannot represent the entangled state $\tilde{\Psi}$.
To represent $\tilde{\Psi}$ using Eq.~\eqref{eq:cas22_ducc_inexact}, we must have $t_1 = t'_1$; however, according to Eq.~\eqref{eq:toyproblem_ducc_inexact} the magnitude of $c_2$ (and $c_3$) is bound by the inequality
\begin{equation}
\label{eq:toyproblem_inequality}
|c_2| = |\cos(t_1) \sin(t_1)| = \frac{1}{2} |\sin(2 t_1)| \leq \frac{1}{2}.
\end{equation}
As a consequence, the state $\tilde{\Psi}$ cannot be represented with this \theoryname ansatz [Eq.~\eqref{eq:toyproblem_ducc_inexact}] since the coefficients $c_2$ and $c_3$ fall outside the bound imposed by Eq.~\eqref{eq:toyproblem_inequality}.

We may also analyze this second \theoryname ansatz by considering the amplitudes as a function of $s$, in an analogy with the path formalism developed for UCC.
Critical points correspond to the zeros of the determinant
\begin{equation}
\det \mathbf{M} = \frac{1}{4} [\cos(2 t_1) + \cos(2 t'_1)]^2,
\end{equation}
which are depicted in Fig.~\ref{fig:critical} C.
If we try to connect the reference $\Phi_0$ to the state $\tilde{\Psi}$ with a path $\Psi(s)$ that is constrained to satisfy $c_2 = c_3$, we find that when the amplitudes reach the critical values $t_1= t'_1 = \pi/4$ then the state is represented by the coefficient vector
\begin{equation} 
\frac{1}{2}
\begin{pmatrix}
\cos t_2  - \sin t_2 \\ 1  \\ 1 \\ \cos t_2  + \sin t_2
\end{pmatrix}.
\end{equation}
The gradients with respect to $t_1$ and $t_2$ are linearly dependent and correspond to
\begin{equation} 
\ket{v_1} \rightarrow 
\begin{pmatrix}
-\cos t_2  - \sin t_2 \\ 0  \\ 0 \\ \cos t_2  - \sin t_2
\end{pmatrix},
\quad
\ket{v_2} = \frac{1}{2} \ket{v_1}.
\end{equation}
In this case, either gradient has zero component along the state $\tilde{\Psi}$, i.e. $\braket{\tilde{\Psi}|v_1} = \braket{\tilde{\Psi}|v_2} = 0$, \textit{independently of} $t_2$, and so no path can be found in which $t_1$ and $t_2$ are varied simultaneously that crosses the critical set.
As a consequence, the second \theoryname ansatz cannot represent the state $\tilde{\Psi}$.

\section{Discussion}
In this work we have investigated several formal aspects of unitary coupled cluster theory and some of its variants currently of interest in the simulation of many-body systems with quantum computers.
By writing the UCC wave function as an integral along a path in Hilbert space we have been able to express the conditions that make UCC an exact representation of arbitrary states in terms of the properties of the set of critical points.
Dimensionality arguments show that the set of critical point has measure zero, which combined with the flexibility of choosing different paths, suggest that the \uccf is likely to be integrable for almost all choices of determinant reference and final state, and thus, in practical usage, exact.
Our numerical experiments confirm this picture and shows that although critical points do exist, they do not prevent integration of the \uccf.
Although the UCC representation of a state is not unique,
in our numerical examples, we find a set of principal solutions characterized by amplitudes defined in the range $t_\mu \in [-\pi,\pi]$, for all $\mu$.

Our second result is a proof that any state may be generated by a \theorypre (factorized) form of UCC.
This representation employs only the particle-hole excitation/de-excitation operators and contains exactly $N_\mathrm{FCI} - 1$ such rotations.
This representation is not unique, in the sense that among the $(N_\mathrm{FCI} - 1)!$ possible sequences of unitary operators, only a subset is exact.
For a simple toy model, we construct two \theoryname wave function, one which is exact and the other one not.
From the exactness of \theoryname, we show that any state may be approximated with arbitrary precision by a product of \textit{particle-hole} one- and two-body unitary transformations acting on a single Slater determinant.
Thus it is not necessary in principle to consider unitary rotations generated by \textit{general} one- and two-body operators.

Our work elucidates several new aspects of UCC theory and produced new formal tools to understand many-body wave functions.
For example, the differential formalism used to analyze UCC is also a practical approach to find the UCC representation of any state (cluster analysis).
Our result on the exactness of \theoryname provides a way to explicitly construct minimum length product unitary transformations that are exact for a given number of electrons.
The ansatz built from an infinite product of particle-hole singles and doubles considered in this work may be an interesting variational ansatz for quantum computing.
This ansatz requires, in principle, an infinite number of terms.
Therefore, an interesting question is whether or not it is possible to represent a state using a product of one- and two-body general or particle-hole unitary rotations such that the number of parameters is equal to the size of the Hilbert space.
Recently there has been some work exploring this direction.\cite{Gard:2019vd}

In practical applications, the ans\"{a}tze considered here must be approximated to reduce the number of variational parameters to a low-order polynomial of the number of electrons.
Then an important question is: What approximations maximize the efficiency with which these ans\"{a}tze represent quantum states relevant to problems in chemistry and condensed matter physics?
Therefore, it would be desirable to perform a thorough numerical comparison of these schemes and related approximate variants in applications to challenging  strongly correlated states.

\begin{acknowledgments}
The authors were supported by the U.S. Department of Energy under Award No.  DE-SC0019374. This work has benefitted from helpful discussions with Dominika Zgid, Nicholas Mayhall, Edwin Barnes, and Piotr Piecuch.
\end{acknowledgments}

\bibliography{papers3.bib,special.bib}

\end{document}